\documentclass[11pt, twoside, a4paper]{article}

\title{Emulsification in binary liquids containing colloidal particles: a structure-factor analysis}

\author{Job H J Thijssen$^{1,2*}$ and Paul S Clegg$^{1,2}$}

\newcommand{\degC}{$^{\circ} \mathrm{C}$}
\newcommand{\degCpm}{$^{\circ} \mathrm{C} \cdot \mathrm{min}^{-1}$}
\newcommand{\fd}{\mathrm{d}}
\newcommand{\micron}{$\mu \mathrm{m}$}
\newcommand{\Rc}{$R_{\mathrm{c}}$}
\newcommand{\Sq}{$S \left( q \right)$}
\newcommand{\PDD}{$\mathrm{PD}_{\mathrm{2D}}$}

\usepackage[dvips]{graphicx}

\begin{document}

\maketitle

\noindent{\footnotesize{\textit{$^{1}$ SUPA School of Physics and Astronomy, The University of Edinburgh, Edinburgh EH9 3JZ, United Kingdom}}}

\noindent{\footnotesize{\textit{$^{2}$ COSMIC, The University of Edinburgh, Edinburgh EH9 3JZ, United Kingdom}}}

\noindent{\footnotesize{$^{*}$ j.h.j.thijssen@ed.ac.uk}}

\bigskip


\smallskip


\begin{abstract}
We present a quantitative confocal-microscopy study of the transient and final microstructure of particle-stabilised emulsions formed via demixing in a binary liquid. To this end, we have developed an image-analysis method that relies on structure factors obtained from discrete Fourier transforms of individual frames in confocal image sequences. Radially averaging the squared modulus of these Fourier transforms \emph{before} peak fitting allows extraction of dominant length scales over the entire temperature range of the quench. Our procedure even yields information just after droplet nucleation, when the (fluorescence) contrast between the two separating phases is scarcely discernable in the images. We find that our emulsions are stabilised on experimental time scales by interfacial particles and that they are likely to have bimodal droplet-size distributions. We attribute the latter to coalescence together with creaming being the main coarsening mechanism during the late stages of emulsification and we support this claim with (direct) confocal-microscopy observations. In addition, our results imply that the observed droplets emerge from particle-promoted nucleation, possibly followed by a free-growth regime. Finally, we argue that creaming strongly affects droplet growth during the early stages of emulsification. Future investigations could clarify the link between quench conditions and resulting microstructure, paving the way for tailor-made particle-stabilised emulsions from binary liquids.
\end{abstract}

\section{Introduction}\label{sec:intro}

Particle-stabilised emulsions, in which colloidal particles rather than molecular surfactants provide stabilisation vs macroscopic phase separation, have received increasing attention over the past two decades \cite{BinksHorozov2008}. The reasons for this growing interest have been twofold: 1) particle-stabilised emulsions are model arrested systems and 2) they have significant potential for applications in the food, personal-care, pharmaceutical, agricultural and petrochemical industries \cite{BinksHorozov2008, Hunter2008, LealCalderon2008}. Typically, these so-called Pickering-Ramsden emulsions are fabricated by direct mixing of two immiscible liquids in the presence of colloidal particles, e.g.~via limited coalescence \cite{Arditty2003}. In this process, an excess of liquid-liquid interface is created, i.e.~there are not enough particles to cover all of it. Subsequently, the emulsion coarsens, thereby reducing the surface-to-volume ratio of the droplets, until the area fraction of interfacial particles is sufficient to stabilise them. Allowing superior control over droplet size and variance, microfluidic techniques have also been employed to fabricate Pickering-Ramsden emulsions \cite{Nie2008}. Recently, using partially miscible rather than immiscible liquids, a promising (reversible) alternative for the fabrication of particle-stabilised emulsions has been reported --- demixing via nucleation and growth of droplets in a binary liquid containing colloidal particles \cite{Clegg2007}.

\begin{figure}[!h]
\begin{center}
\includegraphics[width=0.99\textwidth]{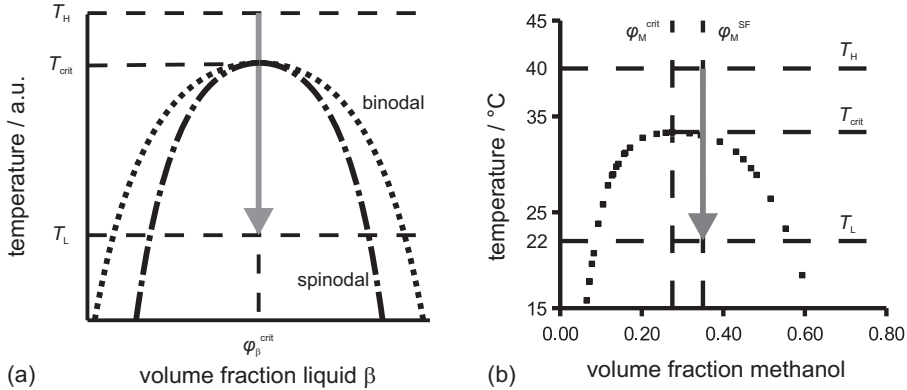}
\caption{(a) Schematic phase diagram of a symmetric binary liquid $\alpha \beta$: $\varphi_{\beta}^{\mathrm{crit}}$ is the critical volume fraction of liquid $\beta$. Grey arrow: a deep quench from the single-fluid phase into the unstable region of the phase diagram, inducing spinodal decomposition. (b) Phase diagram for hexane-methanol \cite{Hradetzky1991}: $\varphi_{\mathrm{M}}^{\mathrm{SF}}$ is the single-fluid volume fraction of methanol; mole fractions were converted into volume fractions using linear fits to the temperature-dependent densities of hexane and methanol \cite{HandbookChemPhys20092010}. Grey arrow: an off-critical quench, used here to form particle-stabilised emulsions via nucleation and growth of droplets. Panel (b) adapted by permission of The Royal Society of Chemistry \cite{Thijssen2010}.}\label{fig:Figure_PhaseDiagram}
\end{center}
\end{figure}

Upon quenching a binary mixture from the single-fluid phase into the two-phase region of its phase diagram (figure \ref{fig:Figure_PhaseDiagram}) \cite{Hradetzky1991, HandbookChemPhys20092010, Thijssen2010}, there are several ways in which the liquid components can separate. If the phase diagram is symmetric, a deep quench through the critical point will induce spinodal decomposition (figure \ref{fig:Figure_PhaseDiagram}(a)): the liquids separate via coarsening of a bicontinuous domain pattern. A shallow, off-critical quench can take the system into the metastable region of the phase diagram, in between the binodal and spinodal lines (figure \ref{fig:Figure_PhaseDiagram}), where demixing proceeds via nucleation and growth of droplets. If the dynamics of the two phases operate at very different time scales, demixing may proceed via viscoelastic phase separation, which can lead to the formation of various (transient) domain patterns \cite{Tanaka2000}.

Domain coarsening is driven by the energy cost of the liquid-liquid interface, which is quantified by the interfacial tension $\gamma$. Interfacial particles provide an alternative means to reduce this (energy-expensive) interfacial area. This is described by
\begin{equation}\label{eq:Uattach}
    \Delta G_{\mathrm{d}} = \pi r^{2} \gamma \left( 1 - \left| \cos{\theta} \right| \right)^{2} \ ,
\end{equation}
where $\Delta G_{\mathrm{d}}$ is the free energy of detachment of a spherical particle of radius $r$ and contact angle $\theta$, as measured through the more polar phase \cite{BinksHorozov2008}. Even for the partially miscible liquids hexane and methanol, which have a relatively low interfacial tension $\gamma$ of order 0.3 $\mathrm{mN} \cdot \mathrm{m}^{-1}$ at room temperature ($T_{\mathrm{room}}$), equation (\ref{eq:Uattach}) predicts $\Delta G_{\mathrm{d}} \sim 6 \times 10^{4} \ k_{\mathrm{B}} T_{\mathrm{room}}$ for an `average' colloidal particle, i.e.~1 \micron{} in diameter \cite{HandbookChemPhys20092010, Abbas1997}. In other words, colloidal particles can become irreversibly attached to liquid-liquid interfaces. This is in contrast with molecular surfactants, which adsorb to and desorb from the interface in a dynamic equilibrium with an excess of surfactant in one of the liquid phases. Hence, in the case of emulsification in the presence of colloidal particles, coarsening will proceed until the surface area of the droplets can only just accommodate all of the particles. The jammed particles then halt further phase separation, because the energy barrier preventing their expulsion from the liquid-liquid interface is too high.

Though not extensively, the microstructure of particle-stabilised emulsions has been studied quantitatively, in the case of fabrication by direct mixing. Arditty \emph{et al.} considered dense emulsions of mm-scale droplets, formed via limited coalescence and stabilised by clusters of silica nanoparticles \cite{Arditty2003}. Note that these samples did not have a surplus of colloids in the continuous phase --- all particles were interfacial from the start of limited coalescence onwards. Using video observations, they found that the average droplet diameter rapidly increases initially, after which it saturates at a limiting value. The corresponding size distributions are monomodal and surprisingly narrow, which is attributed to coarsening via coalescence. In short, small droplets coalesce fast due to their relatively large surface-to-volume ratio (low particle coverage), while large droplets coalesce slowly because film thinning, shape relaxation and local rearrangements occur within time scales that increase with droplet size. Combined, these effects limit droplet polydispersity in their emulsions to 10--20\%.

More recently, Schelero \emph{et al.} considered direct-mixing emulsions in which the micron-sized droplets were stabilised by patches of catanionic crystals \cite{Schelero2009}. Ageing behaviour at room temperature was characterised using single-particle light scattering over a period of nine months. Regardless of sample composition and age, they found droplet-size distributions with three distinctive peaks, which they attribute to 1) pure catanionic crystals, 2) oil droplets covered with a monolayer of catanionic pairs and 3) oil droplets with larger catanionic crystals in addition to the monolayer. Note, however, that their emulsion preparation procedure differs significantly from that of Arditty \emph{et al.}~\cite{Arditty2003}. In particular, Schelero \emph{et al.} added two kinds of surfactant that form crystals upon cooling just \emph{after} mixing all of the ingredients at an elevated temperature. It is not clear to what extent the properties of these emulsions are influenced by 1) the presence of surfactants during emulsification and 2) the lack of solid particles at the start of macroscopic phase separation.

In this paper, we report a quantitative confocal-microscopy study of emulsification upon demixing in a binary liquid containing colloidal particles. To the best of our knowledge, the microstructure of particle-stabilised emulsions prepared by this route has not yet been characterised quantitatively. As emulsification is induced through a controlled temperature quench, the process can be imaged in situ and in real time. Obtaining droplet-size distributions remains challenging, especially near droplet nucleation, where the two demixing liquids are still chemically similar, resulting in an inherently low (fluorescence) imaging contrast. Hence, we have developed an image-analysis method that relies on structure factors obtained from discrete Fourier transforms of individual frames in confocal image sequences. Radially averaging the squared modulus of these Fourier transforms \emph{before} peak fitting allows extraction of dominant length scales over the entire temperature range of the quench, even when droplets are not yet discernable in the images, i.e.~just after nucleation. Thus analysed, our confocal-microscopy data implies that our final emulsions have bimodal droplet-size distributions.

The rest of this paper is organised as follows. Below, in section \ref{sec:methods}, we start with a description of our experimental procedures, notably the image-analysis method that we have developed. We continue by presenting our main results in section \ref{sec:results}, focussing first on the final emulsions and then on emulsification. In section \ref{sec:disc}, we first provide an explanation for the observed bimodal droplet-size distributions, after which we discuss emulsion formation and stability. Finally, in section \ref{sec:conc}, we draw conclusions and suggest future work.

\section{Materials and methods}\label{sec:methods}

Sample preparation and characterisation (sections \ref{subsec:sampleprep}--\ref{subsec:samplecharacterization}) follow methods developed in reference \cite{Thijssen2010}.

\subsection{Sample preparation}\label{subsec:sampleprep}

The experiments were performed with silica colloids in the binary mixture hexane-methanol \cite{Hradetzky1991}. Hydrophobic, fumed silica particles were used as received (Degussa, AEROSIL R812). This aerosil consists of roughly spherical primary particles with an average size $\sim 7$ nm, which are clustered together into $0.4$ \micron{} diameter fractal aggregates; these aggregates, as a rule, cannot be broken down further \cite{Degussa1993, Iannacchione1998}. Hexane (Fluka, $\ge 99$\%), methanol (Fisher Scientific, $99.99$\%) and the fluorescent dye Nile Red (Sigma) were used as received. Nile Red was usually dissolved in the methanol at concentrations of $3.7 \cdot 10^{-4}$ to $2.9 \cdot 10^{-3}$ M. Sample mixtures were typically prepared at a hexane/methanol volume ratio of 65/35 (61/39 w/w) and a silica content of 2.0 vol\% (5.4 wt\%). All amounts were determined by weighing. The particles were dispersed using an ultrasonic processor at 6 W for 2 minutes (Sonics, Vibra-cell), with part of the vial immersed in water, followed by 10 s of vortex mixing. Evaporation losses during sonication led to a maximum deviation of 2.5\%-point in the hexane/methanol volume ratio.

\subsection{Sample transfer and cooling}\label{subsec:sampletransfercooling}

Sample cells were rectangular and had a 1 mm internal path length (Starna Scientific Ltd). Sample mixtures in the single-fluid phase were transferred to these cells in an incubator at approximately 40 \degC{} (Stuart, SI60). They were then quickly transferred to a modified hotstage at 40.0 \degC{} (Linkam Scientific LTS350). Emulsification was studied while cooling from 40.0 \degC{} to 22.0 \degC{} at 5.0 \degCpm{} (figure \ref{fig:Figure_PhaseDiagram}(b)). During an experimental session, samples were typically subjected to several heating and cooling cycles. After a heating cycle, they were vigorously shaken by hand in the incubator at $\sim 40$ \degC{} to ensure proper particle re-dispersion. For samples with less than 1.0 vol\% of silica (figure \ref{fig:Figure_Emulsions_Formed}(c) only), we deviated from the standard procedure described above: they were quenched from $\sim 40$ \degC{} by putting them on a metal surface at $\sim 22$ \degC{} \cite{Thijssen2010}.

\subsection{Sample characterisation by confocal microscopy}\label{subsec:samplecharacterization}

The samples were studied with confocal laser scanning microscopy in reflection, fluorescence and/or transmission. A Nikon ECLIPSE E800/TE300 upright/inverted microscope was used in conjunction with a BioRad Radiance 2100 scanning system. The 457 nm line of an Ar-ion laser was used for reflection, while a 543 nm HeNe laser was employed to excite the Nile Red. Emission filters were used as appropriate. Visual inspection and confocal microscopy on samples without particles confirmed that the recorded Nile Red fluorescence was mainly coming from the methanol-rich phase. Owing to the hotstage, a Nikon Plan Fluor Extra Long Working Distance $20\times$/0.45 NA objective with an adjustable correction collar was used.

\subsection{Image analysis}\label{subsec:imageanalysismethods}

For quantitative analysis, confocal fluorescence images were enhanced and analysed using the IDL software package (RSI v6.3). First, to correct for non-uniform brightness, IDL's `adaptive histogram equalization' function was applied to our images. Subsequently, their 2D Fast Fourier Transforms were calculated and the modulus of these 2DFFTs was squared. To simplify the fitting procedure (see below), spatial frequencies close to 0, related to the finite size of the images, were removed using a Butterworth high-pass filter; within each analysed confocal image sequence, the values of the filter parameters were the same for all frames. After that, the $\left| \mathrm{2DFFT} \right|^2$ arrays were radially averaged and the resulting structure factors $S \left( q \right)$ were smoothed using a typical box width of 3 to 5 reciprocal pixels.

Using an iterative Levenberg-Marquardt algorithm, peaks in experimental structure factors $S \left( q \right)$ were fitted to an `Extreme function',
\begin{equation}\label{Eq:Extreme_Function_Formula}
    y = y_{0} + A \cdot \mathrm{e}^{\left( -\mathrm{e}^{-z} - z + 1 \right)} \ ,
\end{equation}
where
\begin{equation}\label{Eq:Extreme_Variable_z}
    z = \frac{x-x_{\mathrm{c}}}{w} \ ,
\end{equation}
$y = S \left( q \right)$, $y_{0}$ is the offset, $A$ is the peak amplitude, $x_{\mathrm{c}}$ is the peak centre and $w$ is its width. Because the images in our confocal time series are 2D confocal slices through a 3D emulsion, we define the polydispersity \PDD{} of the population of cross-sectional disks corresponding to a particular peak in \Sq{}:
\begin{equation}\label{Eq:Polydispersity_Definition}
    \mathrm{PD}_{\mathrm{2D}} = \left( \frac{w}{x_{\mathrm{c}}-w} \right) \cdot 100\% \ ,
\end{equation}
which roughly corresponds to the relative Half Width at Half Maximum of the fitted peaks. The pixel size $\Delta q \times \Delta q$ of the 2DFFT images was calibrated using
\begin{equation}\label{Eq:qcal}
    \Delta q \left[ \mathrm{\mu m^{-1}} \right] = \frac{1}{N \cdot \Delta u \left[ \mu m \right]} \ ,
\end{equation}
where $\Delta u \times \Delta u$ is the pixel size of the corresponding $N \times N$ confocal-microscopy image. Finally, we define the dominant length scale in an image as
\begin{equation}\label{Eq:dominant_length_scale}
    L \left[ \mu m \right] = \frac{1}{x_{\mathrm{c}} \left[ \mathrm{\mu m^{-1}} \right]} \ .
\end{equation}

Peak fitting allowed for batch processing of images in time series of demixing by first analysing the last image $\left( N-1 \right)$, using the resulting parameters as an initial parameter estimate for analysing image $\left( N-2 \right)$ and so on, thus `backtracking' peaks in $S \left( q \right)$ as a function of temperature. While this avoids the need to enter trial parameters for each image, it does result in spurious fits above the nucleation temperature, i.e.~when the sample is in the single-fluid phase.

\section{Results}\label{sec:results}

\subsection{Final emulsions}\label{subsec:emulsions}

\begin{figure}
\begin{center}
\includegraphics[width=0.99\textwidth]{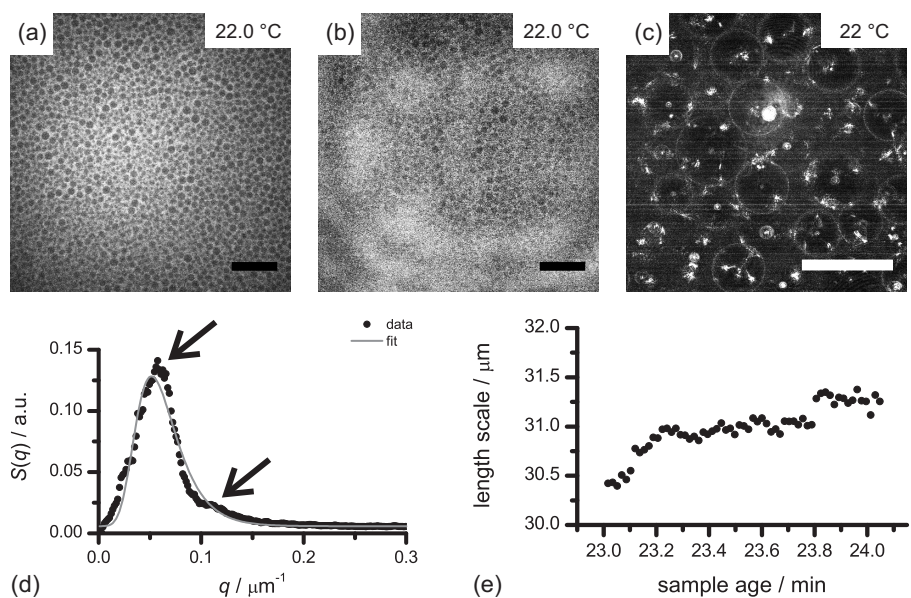}
\caption{(a,b) Confocal images of a hexane-in-methanol emulsion stabilised by 2.0 vol\% of silica: (a/b) fluorescence/reflection reveals the methanol-rich phase/silica (white). Images recorded at 22.0 \degC{}, following a quench from 40.0 \degC{} at 5.0 \degCpm{}. (c) Confocal reflection image showing interfacially trapped particles (0.1 vol\% silica). (d) Structure factor \Sq{} corresponding to (a); black arrows point to \Sq{} peaks. (e) Dominant length scale $L$ versus time in an emulsion, containing 1.0 vol\% of silica particles, at 22.0 \degC{} (\emph{Appendix \ref{sec:movieguide}}). Images were recorded at 14 \micron{} from the top or (c) 0.79 mm from the bottom of the sample. Scale bars: 100 \micron{}. Panel (c) reproduced by permission of The Royal Society of Chemistry \cite{Thijssen2010}.}\label{fig:Figure_Emulsions_Formed}
\end{center}
\end{figure}

To facilitate image analysis later on, we first summarise our previous results concerning the qualitative structure of hexane-methanol/silica emulsions that had been formed by quenching from the single-fluid phase \cite{Thijssen2010}. Figures \ref{fig:Figure_Emulsions_Formed}(a) and (b) show the confocal fluorescence and the corresponding confocal reflection images of such an emulsion, just after the quench target temperature had been reached. The droplets contain the less dense, hexane-rich phase \cite{Abbas1997}, while the continuous phase is rich in methanol. Although the silica particles mainly partition into the continuous phase, the confocal reflection image in figure \ref{fig:Figure_Emulsions_Formed}(c) implies that our emulsions are stabilised by interfacially trapped particles, leaving an excess of colloids in the continuous phase if the particle volume fraction $\gg 0.1$\%. This implication is corroborated by figure \ref{fig:Figure_Emulsions_Formed}(e), as it seems unlikely that colloids in the continuous phase alone can stabilise our emulsions for at least 24 minutes at silica volume fractions as low as 1.0 vol\%.

% jtCHECK: OTR !!!

To complement these qualitative observations, we calculate structure factors $S \left( q \right)$ from confocal fluorescence images, thereby revealing that our emulsions may well be characterised by bimodal droplet-size distributions. Figure \ref{fig:Figure_Emulsions_Formed}(d) shows the structure factor corresponding to figure \ref{fig:Figure_Emulsions_Formed}(a); the sample had just reached the quench target temperature. The main peak at low $q$ is related to the average distance between large cross-sectional disks in the 2D confocal slices of our 3D emulsions; a value of $\left( 19.4 \pm 0.6 \right)$ \micron{} is obtained from the corresponding fit in this case. As the cross-sectional disks are (nearly) close-packed, this should roughly correspond to their average (apparent) diameter. Note that figure \ref{fig:Figure_Emulsions_Formed}(d) presents a `worst-case scenario' --- the width of the main \Sq{} peak is generally fitted well at half height, but the peak asymmetry is usually captured more accurately. Considering the relatively high polydispersity \PDD{} $\approx 62$\% (equation (\ref{Eq:Polydispersity_Definition})), it is unlikely that the second peak at higher $q$ stems from the corresponding form factor \cite{Cumming1990}. Instead, we attribute it to the average distance between two small droplets or between a small and a large one. In either case, its presence implies a bimodal size distribution. The shoulder on the left side of the main peak in figure \ref{fig:Figure_Emulsions_Formed}(d) is probably an artefact of limited resolution --- it may correspond to pairs of neighbouring droplets that appear as dumbbells to the software. Having considered the structure of the final emulsions, we will next use this structure-factor analysis to explore the emulsification process.

\subsection{Emulsification}\label{subsec:emulsification}

\begin{figure}
\begin{center}
\includegraphics[width=0.99\textwidth]{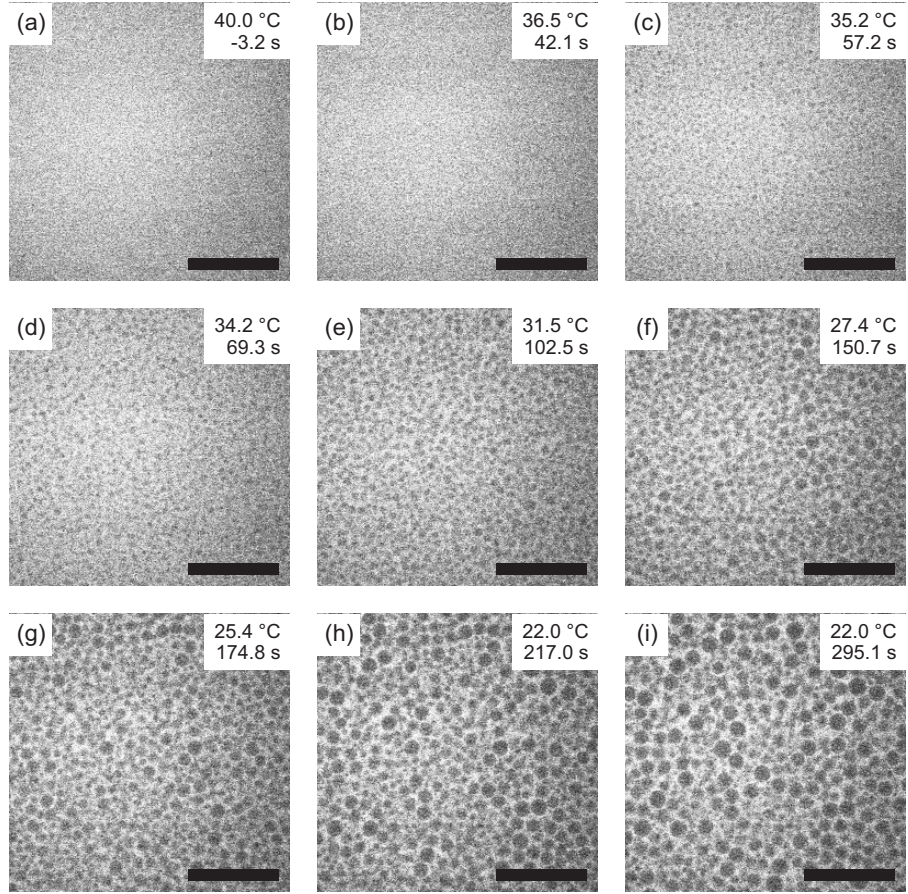}
\caption{Selected $xy$-frames, analysed in figure \ref{fig:Figure_Tracking_Graphs} and centrally cropped here ($512 \times 512 \rightarrow 256 \times 256$ pixels), from a confocal fluorescence image sequence of emulsification in a hexane-methanol mixture containing 2.0 vol\% of silica particles (see \emph{Appendix \ref{sec:movieguide}} for movie and \emph{Appendix \ref{sec:strucfac4imagseries}} for structure factors); panel (h) is a digital zoom of figure \ref{fig:Figure_Emulsions_Formed}(a). Starting at $t = 0$ s, the sample was quenched from 40.0 \degC{} to 22.0 \degC{} at 5.0 \degCpm{}, resulting in the nucleation and growth of hexane-rich droplets in a methanol-rich continuous phase. Images were recorded at 14 \micron{} from the top of the sample. Scale bars: 100 \micron{}.
}\label{fig:Figure_Image_Series}
\end{center}
\end{figure}

% jtCHECK: OTR !!!
Confocal fluorescence image sequences of the emulsification process provide a clue as to the origin of the observed bimodal droplet-size distributions. Figure \ref{fig:Figure_Image_Series} shows a snapshot montage extracted from such a sequence (\emph{Appendix \ref{sec:movieguide}}). At the start of the quench (figure \ref{fig:Figure_Image_Series}(a)), the fluorescence signal is flat across the frame, which is consistent with a binary-liquid/particle system in its single-fluid phase. At the nucleation temperature (figure \ref{fig:Figure_Image_Series}(b)), as determined from an intensity drop in the transmission channel, the fluorescence image is still flat, probably because 1) our confocal resolution is not sufficient to resolve the nucleated droplets \cite{Baumberger1992} and 2) the two demixing liquids still have similar chemical compositions, limiting the (fluorescence) contrast. Once droplets have become discernable (figures \ref{fig:Figure_Image_Series}(c) and (d)), phase separation appears to have yielded a monomodal size distribution. Upon further cooling, however, larger droplets start popping up in between consecutive confocal frames (figures \ref{fig:Figure_Image_Series}(f)--(i)). We attribute this to coarsening, at this stage, being dominated by coalescence together with creaming, clearly resulting in a multimodal size distribution.

\begin{figure}
\begin{center}
\includegraphics[width=0.5\textwidth]{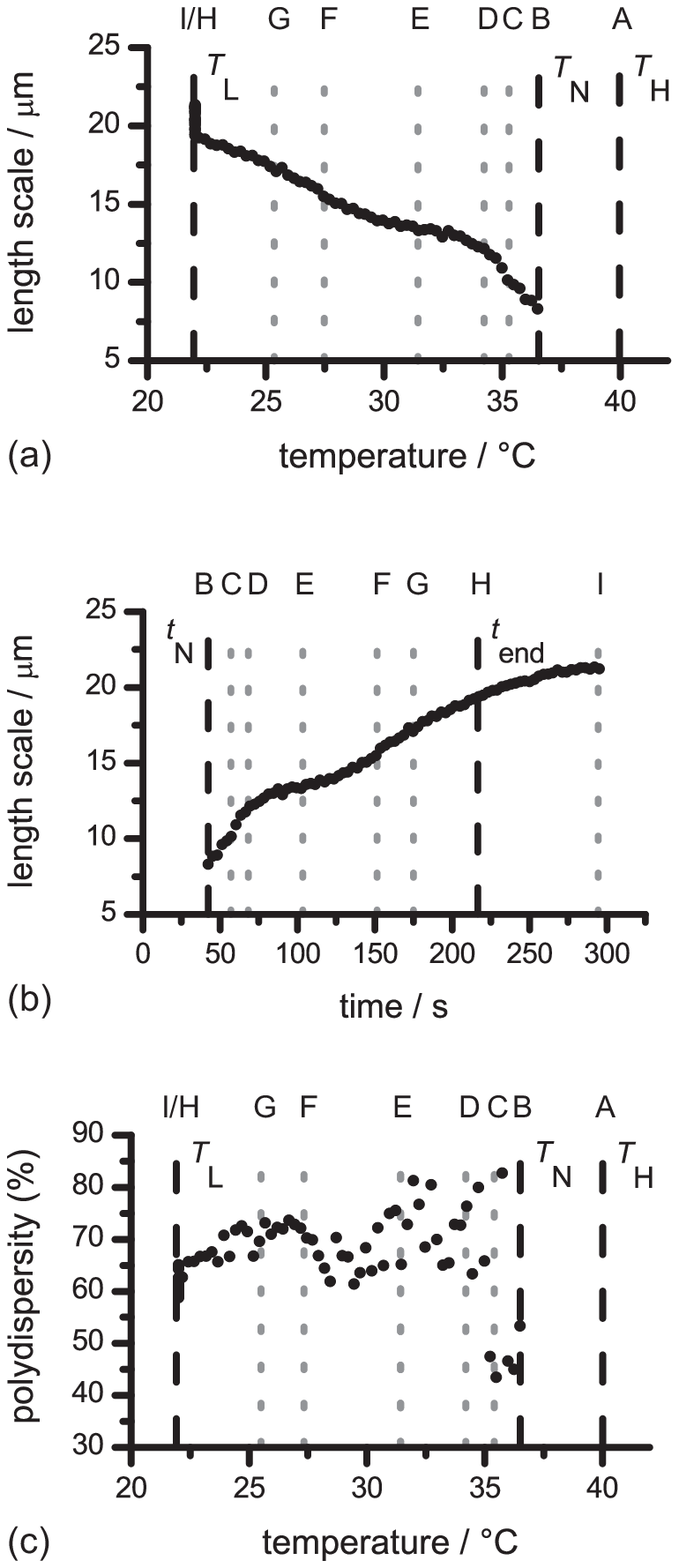}
\caption{Probing the dominant length scale in the image sequence shown in figure \ref{fig:Figure_Image_Series}; the uppercase letters `A' to `I' correspond to the lowercase panel labels in figure \ref{fig:Figure_Image_Series}. Note that the images in figure \ref{fig:Figure_Image_Series} have been centrally cropped for clarity ($512 \times 512 \rightarrow 256 \times 256$ pixels), but the original images have been analysed for the graphs above. (a,b) Dominant length scale and (c) polydispersity \PDD{} as a function of (a,c) temperature and (b) time.
}\label{fig:Figure_Tracking_Graphs}
\end{center}
\end{figure}

To quantify this coarsening behaviour, experimental structure factors were fitted to a functional form (equations (\ref{Eq:Extreme_Function_Formula}) and (\ref{Eq:Extreme_Variable_z})), thereby allowing batch processing of confocal images. Figure \ref{fig:Figure_Tracking_Graphs}(a) shows a graph of the dominant length scale $L$ as a function of temperature (equation (\ref{Eq:dominant_length_scale})), the features of which can be explained by comparing it to the corresponding image sequence in figure \ref{fig:Figure_Image_Series}. Before droplet nucleation (A--B), the confocal images contain no dominant length scale, apart from the image size and the pixel size. At the nucleation temperature (B), determined independently from a fall in transmission intensity, our software picks up a reasonable length scale, even though no droplets are discernable in the corresponding confocal image (figure \ref{fig:Figure_Image_Series}(b))! Just after droplet nucleation (B--D), $L$ increases quickly as the sample is quenched deeper into the two-phase part of the phase diagram. From (D--E), the growth rate falls, but it picks up again at temperatures below (E),\footnote{In similar graphs for emulsions containing 2.5 and 3.5 vol\% of silica (not shown here), the increase in the rate of droplet growth at temperatures below (E) is less pronounced (or absent).} which we ascribe to coalescence taking over as the main coarsening mechanism (figure \ref{fig:Figure_Image_Series}(e--i)).

As indicated by the vertical line at 22.0 \degC{} in the graph in figure \ref{fig:Figure_Tracking_Graphs}(a), our emulsions do not necessarily stop coarsening immediately after the quench. This is revealed more clearly in figure \ref{fig:Figure_Tracking_Graphs}(b), in which the same data has been re-plotted as a function of time. For this sample, just after the target temperature was reached, the coarsening rate started to decrease. However, it took about a minute before the dominant length scale reached a plateau value. Judging from experiments with a thermocouple embedded in a sample without colloids, the time lag between sample and hotstage temperatures during a 5.0 \degCpm{} quench may be as large as 50 s in our set-up, which could explain the residual coarsening. However, in similar graphs for emulsions containing 2.5 and 3.5 vol\% of silica (not shown here), the dominant length scale does not reach a plateau value within the same time interval, so it is unclear whether our emulsions are stable immediately after the quench. To probe the stability of our emulsions on a longer time scale, the dominant length scale $L$ in a 23-minute old emulsion was extracted from a one-minute time series at constant temperature (figure \ref{fig:Figure_Emulsions_Formed}(e)). This graph demonstrates that our emulsions are stable on experimental time scales, for $L$ is virtually constant; the two jumps near $t = 23.2$ and $23.8$ min correspond to re-arrangements coinciding with coalescence events (\emph{Appendices \ref{sec:imagcoal} and \ref{sec:movieguide}}).

\section{Discussion}\label{sec:disc}

\subsection{Emulsion structure}\label{subsec:emulstrucdisc}

\begin{figure}
\begin{center}
\includegraphics[width=0.99\textwidth]{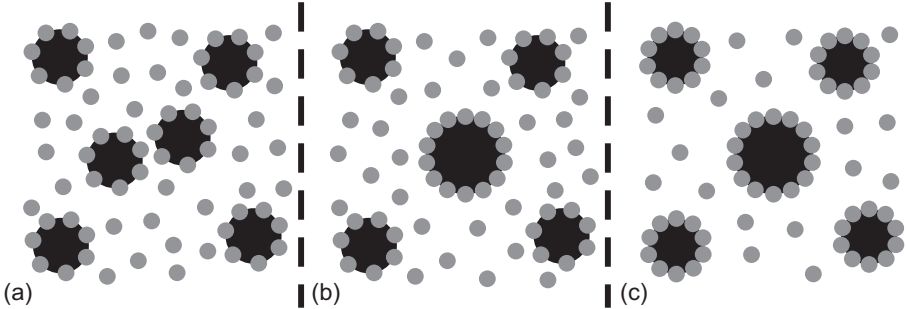}
\caption{2D schematic explaining bimodal droplet-size distributions in emulsions formed upon quenching a binary-liquid/particle mixture from its single-fluid phase. (a) Just after droplet nucleation, the droplets (black) are only partly covered with particles (grey). (b) When two droplets merge, the resulting droplet has a smaller surface-to-volume ratio (equation (\ref{eq:surf2volratio})), i.e.~it has a denser particle coating and is thus more stable. (c) Droplets that have not merged can sweep up additional particles from the continuous phase (white), thus stabilising themselves vs coalescence. This emulsification process results in two `generations' of stable droplets with different average sizes.}\label{fig:Figure_Schematic}
\end{center}
\end{figure}

The two peaks in the structure factor in figure \ref{fig:Figure_Emulsions_Formed}(d) imply that our emulsification route leads to bimodal particle-stabilised emulsions, which we have attributed to coalescence together with creaming (figure \ref{fig:Figure_Image_Series}). The underlying mechanism we propose is schematically depicted in figure \ref{fig:Figure_Schematic}. Early in the emulsification process, the liquid-liquid interface is not fully coated with particles (figure \ref{fig:Figure_Schematic}(a)). Though there is a surplus of colloids in the continuous phase (figure \ref{fig:Figure_Emulsions_Formed}(b)), the droplets are mainly stabilised by interfacial particles (figure \ref{fig:Figure_Emulsions_Formed}(c)), so a partial coating is not sufficient to prevent them from merging. As volume is conserved upon coalescence, the surface-to-volume ratio $\nu$ decreases when two spherical droplets of radii $R_{1}$ and $R_{2}$ merge to form a spherical droplet of radius $R_{3}$ (figure \ref{fig:Figure_Schematic}(a) $\rightarrow$ \ref{fig:Figure_Schematic}(b)):
\begin{equation}\label{eq:surf2volratio}
\begin{array}{lll}
\nu_{1} & = \frac{4 \pi R_{1}^2}{\left( 4/3 \right) \pi R_{1}^3} & = \frac{3}{R_{1}} \\
\nu_{3} & = \left( \frac{3}{R_{1}} \right) \cdot \left( \frac{1}{\sqrt[3]{1 + \left( \frac{R_{2}}{R_{1}} \right)^3}} \right) & < \nu_{1} \ .
\end{array}
\end{equation}
Since the number of interfacial particles is (approximately) conserved during coalescence, their area fraction increases, which yields droplets with more complete particle coatings, i.e.~a more stable emulsion (figure \ref{fig:Figure_Schematic}(b)). As the quench progresses, unmerged droplets can sweep up more particles from the surplus in the continuous phase, thereby completing their coatings (figure \ref{fig:Figure_Schematic}(c)). Note that this last step does not occur during limited coalescence, for all particles are interfacial in that process. This may very well be why Arditty \emph{et al.} reported monodisperse emulsions \cite{Arditty2003}, whereas we observe two `generations' of droplets with different average diameters.

At this point, it is worthwhile recalling that our confocal images are 2D slices at a distance $d = 14$ \micron{} from the top of the sample and, hence, they do not necessarily represent the whole 3D emulsion. However, we will argue below that, in our case, a bimodal size distribution for the 2D cross-sectional disks implies a bimodal size distribution for the corresponding 3D droplets, even though the \emph{exact} shapes of their graphs may differ. First of all, our emulsions appear to be random packings of polydisperse spherical droplets in 3D \cite{Thijssen2010}. It is unlikely that a 2D slice through such a 3D emulsion would yield a bimodal disk-size distribution from a monomodal droplet-size distribution. Secondly, assuming that all droplets in the imaging plane have a 3D radius $R > d/2$ and are touching the sample-cell wall, which is not unreasonable during the late stages of emulsification in which a compact cream has formed at the top of the sample, one can write down the relation between $R$ and the 2D disk radius $r$:
\begin{equation}\label{eq:2Dto3Dradii}
    R = \frac{r^2 + d^2}{2d} \ .
\end{equation}
As $r \ge 0$, equation (\ref{eq:2Dto3Dradii}) expresses a one-to-one correspondence between $r$ and $R$, i.e.~if the 2D disks are characterised by a bimodal size distribution, then so are the 3D droplets. Finally, 3D simulations on nucleation and growth of droplets in similar systems have also found bimodal droplet-size distributions which can be ascribed to coalescence \cite{Harting2010}; our experimental results seem to confirm these simulations.

\subsection{Emulsification}\label{subsec:emulsificationmechanism}

Our emulsification process begins with the nucleation of hexane-rich droplets upon cooling the sample from the single-fluid phase into the two-phase region of the phase diagram at an off-critical composition (figures \ref{fig:Figure_Image_Series}(b/c) and \ref{fig:Figure_PhaseDiagram}(b)). Intriguingly, nucleation may happen at temperatures above the binary-liquid binodal, at 36.5 \degC{} instead of 33.2 \degC{} in this case (figures \ref{fig:Figure_Tracking_Graphs}(a) and \ref{fig:Figure_PhaseDiagram}(b)) \cite{Hradetzky1991}, though the actual value of the nucleation temperature varies from sample to sample. This shift is partly due to a temperature lag between sample and hotstage, which can be as large as 0.5 \degC{} in our set-up. In addition, contamination with only a few vol\% of water, which is easily picked up from the air, increases the consolute temperature of the binary mixture by several \degC{} \cite{Alessi1989}.\footnote{Contamination with water can be minimised by storing the hexane over molecular sieves and the methanol under dry nitrogen gas until the day of the experiment.} Finally, colloidal particles may raise the binodal of a binary liquid, by as much as 1.0 \degC/vol\% in this system \cite{Thijssen2010}.

An elevated nucleation temperature that varies with particle volume fraction is likely to imply heterogeneous nucleation \cite{Penner2009}. Indeed, it is essentially impossible to prevent heterogenous nucleation when quenching off-critical mixtures \cite{Baumberger1992}. However, given the strong preference of our silica particles for the continuous methanol-rich phase, it seems unlikely that the hexane-rich phase would nucleate onto these colloids to form droplets. Instead, we propose that the solid particles acquire a methanol-rich layer due to adsorption/wetting, thereby partially depleting the volume in between the particles of methanol, i.e.~locally increasing hexane supersaturation. This would indeed result in particle-promoted nucleation of hexane-rich droplets at temperatures above the binodal of the binary liquid itself.

During the initial stage of emulsification (B-D), liquid-liquid demixing appears to yield a monomodal droplet-size distribution. Though merely speculation, this could be related to the inherently narrow size distribution resulting from free growth of droplets in a dilute emulsion with a continuous phase at quasi-constant supersaturation (\emph{Appendix \ref{sec:freegrowthmono}}). Indeed, figure \ref{fig:Figure_Image_Series}(c/d) shows that our emulsions are initially quite dilute. The corresponding volume fraction of droplets ($\phi_{\mathrm{droplet}}$) can be estimated using the lever rule,
\begin{equation}\label{eq:leverrule}
    \phi_{\mathrm{droplet}} = \left| \frac{\varphi_{\mathrm{M}}^{\mathrm{SF}} - \varphi_{\mathrm{M}}^{\mathrm{MR}}}{\varphi_{\mathrm{M}}^{\mathrm{MR}} - \varphi_{\mathrm{M}}^{\mathrm{HR}}} \right| \ ,
\end{equation}
where $\varphi_{\mathrm{M}}^{\mathrm{SF}}$ is the volume fraction of methanol in the single-fluid phase and $\varphi_{\mathrm{M}}^{\mathrm{HR(MR)}}$ is the volume fraction of methanol in the hexane(methanol)-rich phase \cite{Chaikin2003}. At the same temperature difference from the binodal as in figure \ref{fig:Figure_Image_Series}(d), equation (\ref{eq:leverrule}) predicts $\phi_{\mathrm{droplet}} \sim 0.32$, which is well below the onset of crystallisation for monodisperse hard spheres \cite{Schaertl1994}. In addition, we expect the methanol-rich continuous phase to be constantly supersaturated with hexane, as our quench is continuous. However, given a $1.20 \ \mu\mathrm{m} \times 1.20 \ \mu\mathrm{m}$ pixel size, a critical nucleation radius $\sim 0.03$ \micron{} \cite{Baumberger1992, Fletcher1958} and a particle-cluster size $\sim 0.4$ \micron{} (section \ref{subsec:sampleprep}), it is unlikely that we have probed the free-growth regime itself --- we may have observed a remnant thereof.

During the intermediate stage of emulsification (D--E), droplet growth in the imaging plane, which is near the top of the sample, slows down (figure \ref{fig:Figure_Tracking_Graphs}(a)). This may be explained by the disappearance of (three) buoyancy-related contributions to droplet growth as the cream develops. Firstly, part of the growth in the initial stage may have been caused by large droplets rising into the imaging plane by steady creaming \cite{Herzig2009}, an effect that will contribute less and less as the observed droplet layer densifies. Secondly, as the cream becomes more compact, the local volume fraction of methanol decreases, i.e.~the system moves towards the hexane-rich branch of the phase diagram (figure \ref{fig:Figure_PhaseDiagram}(b)) \cite{Thijssen2010}. This means that, in/near the imaging plane, the hexane supersaturation of the continuous phase decreases, which slows down droplet growth. Finally, creaming will induce convective flows that may contribute to droplet growth at least as much as diffusion does \cite{Abbas1997, Baumberger1992}.

During the final stages of emulsification, at temperatures below (E), coalescence takes over as the main coarsening mechanism (figure \ref{fig:Figure_Image_Series} and \emph{Appendices \ref{sec:imagcoal}} and \emph{\ref{sec:movieguide}}). The reasons for this are twofold. First of all, buoyancy has resulted in a compact cream at the top of the sample, in which many droplets are now close enough to merge. Secondly, as the temperature has now significantly fallen below the binodal, the interfacial tension between the two liquid phases has increased substantially \cite{Abbas1997}. This drives emulsion coarsening through coalescence as the system attempts to reduce the total free-energy cost associated with liquid-liquid contact area. In addition, increasing interfacial tension promotes particle attachment (equation (\ref{eq:Uattach})). As explained in section \ref{subsec:emulstrucdisc}, the presence of colloidal particles on the surface of the droplets eventually stabilises them, resulting in bimodal droplet-size distributions (figure \ref{fig:Figure_Schematic}).

\subsection{Emulsion stability}\label{subsec:emulsionstability}

Particle-stabilised emulsions can be long lived \cite{BinksHorozov2008}; so why do our emulsions slowly coarsen after the quench has reached its target temperature (figures \ref{fig:Figure_Tracking_Graphs}(b) and \ref{fig:Figure_Emulsions_Formed}(e))? Our system is comprised of droplets stabilised by interfacial particles (figure \ref{fig:Figure_Emulsions_Formed}(c)). These fumed silica colloids are fractal clusters of $\sim 7$ nm primary particles and have a total diameter $\sim 0.4$ \micron{} \cite{Degussa1993, Iannacchione1998}. As equation (\ref{eq:Uattach}) predicts a maximum free energy of detachment $\Delta G_{\mathrm{d}} \sim 10^{4} \ k_{\mathrm{B}}T_{\mathrm{room}}$ for 0.4 \micron{} particles \cite{HandbookChemPhys20092010, Abbas1997}, it seems unlikely that they could be expelled from the liquid-liquid interface. However, the surplus of particles in the continuous phase suggests that the particle-liquid-liquid wetting angle $\theta$ is some way away from $90^{\circ}$ (figure \ref{fig:Figure_Emulsions_Formed}(b)) \cite{Thijssen2010}. Moreover, as the fumed silica particles are fractal clusters \cite{Degussa1993, Iannacchione1998}, they cover less interfacial area than equally sized spheres do and hence experience a shallower energy well than that described by equation (\ref{eq:Uattach}). Finally, for partially miscible liquids, Ostwald ripening may also be quite severe \cite{Capek2004}. Taken together, these effects may explain the observed slow coarsening.

\section{Conclusions}\label{sec:conc}

In this paper, we have presented a quantitative confocal-microscopy study of emulsification upon demixing in a binary liquid containing colloidal particles. Our analysis method relies on structure factors \Sq{} calculated from individual images in confocal fluorescence time series via discrete 2D Fast Fourier Transforms. Radially averaging the squared modulus of these 2DFFTs \emph{before} peak fitting allows extraction of droplet size and variance over the entire temperature range of the emulsifying quench.

First of all, our investigation implies that our particle-stabilised emulsions are characterised by bimodal droplet-size distributions.\footnote{Preliminary experiments suggest that slower quenches at 0.1 \degCpm{} result in multimodal rather than bimodal particle-stabilised emulsions.} We attribute this to coalescence together with creaming being the main coarsening mechanism during the late stages of emulsification and we support this claim with (direct) confocal-microscopy observations. Secondly, we have quantitatively characterised the emulsification process itself. Our results imply that it starts with particle-promoted nucleation, possibly followed by free growth of droplets. During the intermediate stage, in which we observe the formation of a compact cream near the top of the sample, coarsening slows down. This may be explained by the disappearance of (three) buoyancy-related contributions to droplet growth:
\begin{enumerate}
\item large droplets no longer move into the imaging plane from below;
\item the increasing packing fraction of hexane-rich droplets in the creamed emulsion reduces hexane supersaturation of the methanol-rich continuous phase;
\item fading of convective flows, which may contribute to droplet growth at least as much as diffusion does.
\end{enumerate}
During the late stages of emulsification, at temperatures significantly below the binodal, the close packing and the increased interfacial tension result in emulsion coarsening through coalescence. We have shown that the presence of interfacial particles eventually stabilises the emulsions on experimental time scales ($\sim \mathrm{minutes}$). However, fumed silica tends to be organised in fractal clusters, which are not efficient in covering liquid-liquid interface. This may explain the slow coarsening of our emulsions at constant temperature.

To conclude, phase separation is a promising route towards the controlled formation of particle-stabilised emulsions. The experimental conditions during emulsification can be controllably varied in-situ and in real time. A more detailed study, involving the tracking of individual droplets in microscopy images, should lead to a better understanding of the relation between quench conditions and emulsion structure. As such, binary liquids containing colloidal particles are an ideal model system for studying emulsification and could potentially allow for tailor-made particle-stabilised emulsions.

\section*{Acknowledgments}\label{unsec:ack}

We would like to thank Degussa for providing particles. Furthermore, we are grateful to T.~Lapp and J.~Vollmer for useful discussions. Finally, we acknowledge funding through EPSRC EP/E030173/01.

\clearpage

\appendix

\section{Imaging coalescence}\label{sec:imagcoal}

% jtCHECK: OTR !!!
\begin{figure}[!h]
\begin{center}
\includegraphics[width=0.99\textwidth]{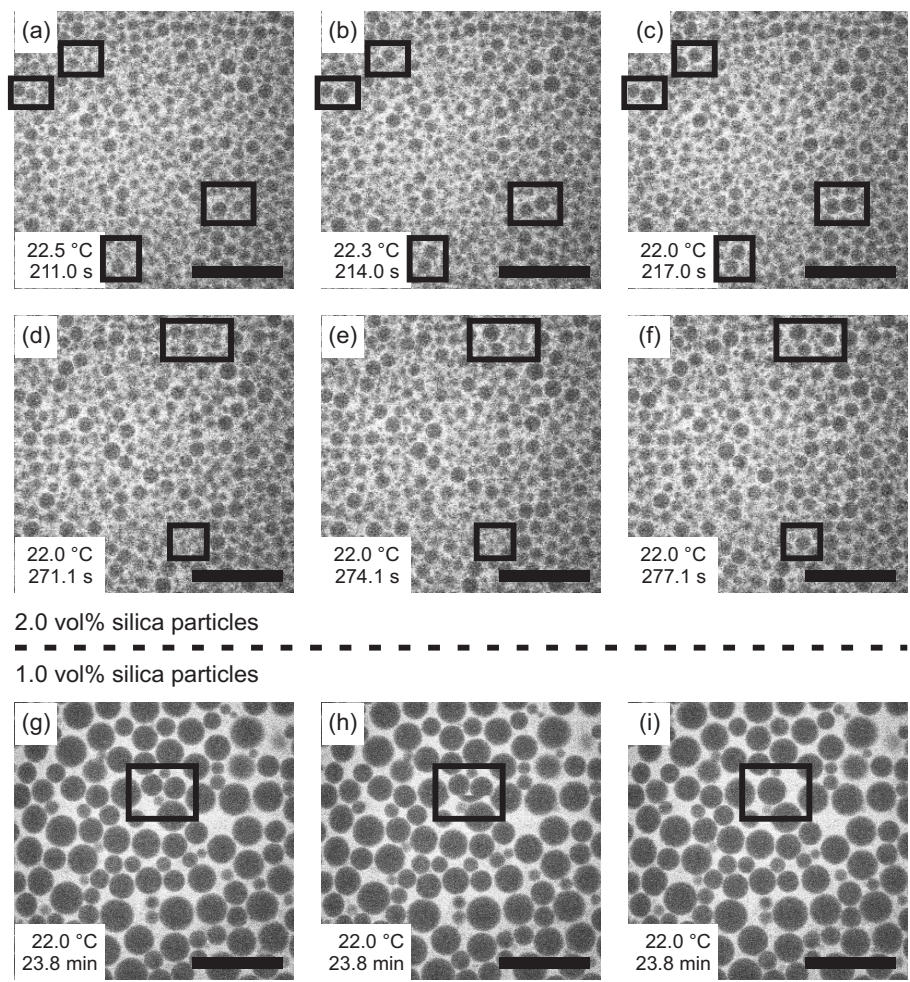}
\caption{Rows: confocal fluorescence image sequences of coalescence events in hexane-methanol emulsions. (a--f) Starting at $t = 0$ s, the sample was quenched from 40.0 \degC{} to 22.0 \degC{} at 5.0 \degCpm{}. These images were taken from the same series as those in figure 3 of the main text. (g--i) The quench target temperature was reached at $t = 0$ s, after which the sample was held at 22.0 \degC{}. These images were taken from the series corresponding to figure 2(e) of the main text (\emph{Appendix \ref{sec:movieguide}}). Images were recorded at 14 \micron{} from the top of the sample. Scale bars: 100 \micron{}.}\label{fig:SI_Figure_Coalescence}
\end{center}
\end{figure}

\clearpage

\section{Monodisperse emulsions from free growth of droplets}\label{sec:freegrowthmono}

To understand why monodisperse, and hence monomodal, emulsions are an inherent feature of droplet nucleation followed by a free-growth regime \cite{Cumming1990, Baumberger1992}, we turn to the growth equation for an isolated droplet in a supersaturated continuous phase
\begin{equation}\label{eq:dropletgrowth}
    \frac{\fd \rho}{\fd \tau} = \rho^{-1} \left( 1 - \rho^{-1} \right) \ ,
\end{equation}
in which
\begin{equation}\label{eq:rhotau}
    \begin{array}{ll}
    \rho & = \frac{R}{R_{\mathrm{c}}} \\
    \tau & = \frac{t}{t_{\mathrm{diff}}} \ .
    \end{array}
\end{equation}
Here, $R$ is the droplet radius, $t$ the time after nucleation and $t_{\mathrm{diff}}$ a diffusion time scale. \Rc{} is the critical radius: nucleated droplets with $R < R_{\mathrm{c}}$ will shrink and disappear, while those with $R > R_{\mathrm{c}}$ will grow. For $R \gg R_{\mathrm{c}}$, equation (\ref{eq:dropletgrowth}) reduces to
\begin{equation}\label{eq:dropletgrowthreduced}
    \rho = \sqrt{2 \tau} \ ,
\end{equation}
\emph{whatever} the initial distribution of droplets \cite{Cumming1990,Baumberger1992}. In words: droplets smaller than \Rc{} disappear in order to reduce the free energy of the system, while droplets much larger than \Rc{} have locally depleted the continuous phase of the nucleating phase, thereby reducing supersaturation and slowing their growth.

Note that it is not unreasonable to assume that equations (\ref{eq:dropletgrowth}) and (\ref{eq:rhotau}) apply in our case as 1) our emulsions are initially quite dilute, 2) the continuous phase is supersaturated (before creaming) and 3) the droplets may very well nucleate in between rather than on the colloidal particles. Even in the case of heterogeneous nucleation, the above argument may hold \cite{Cumming1990, Baumberger1992}, as the critical radius \Rc{} is still a well-defined concept. If one assumes that the embryo nucleating onto a spherical seed particle is itself a portion of a different sphere of radius $R$, then \Rc{} is the critical radius of this embryonic sphere. Notably, the value of \Rc{} is the same as in the case of homogeneous nucleation, even though the volume of the embryo will differ from that of a homogenously nucleated droplet, as all parts of the embryo surface must be in equilibrium with the metastable continuous phase \cite{Fletcher1958}.

\clearpage

\section{Movie guide}\label{sec:movieguide}

See http://tinyurl.com/382tmz2

% jtCHECK: OTR !!!
\begin{itemize}
\item \emph{Thijssen\_2010\_EQA\_SI\_Movie2e\_2xrt.avi}: corresponds to the graph in figure 2(e) of the main text. Hexane-in-methanol emulsion stabilised by 1.0 vol\% of silica particles. The fluorescence reveals the methanol-rich phase (white). At the start of this movie, the sample had already been held at a constant temperature of 22.0 \degC{} for 23 minutes. Images were recorded at a depth of 14 \micron{} from the top of the sample. Image dimensions: $307.97 \ \mu\mathrm{m} \times 307.97 \ \mu\mathrm{m}$. The movie is at $2 \times$ real time; it has been resized from $512 \times 512$ to $384 \times 384$ pixels and subsequently JPEG compressed to reduce file size.
\item \emph{Thijssen\_2010\_EQA\_SI\_Movie3\_6xrt.avi}: corresponds to figure 3 of the main text. Confocal fluorescence image sequence of emulsification in a hexane-methanol mixture containing 2.0 vol\% of silica particles. Starting at $t = 0$ s, the sample was quenched from 40.0 \degC{} to 22.0 \degC{} at 5.0 \degCpm{}, resulting in the nucleation and growth of hexane-rich droplets in a methanol-rich continuous phase. Images were recorded at a depth of 14 \micron{} from the top of the sample and at intervals of 3 s. Image dimensions: $307.97 \ \mu\mathrm{m} \times 307.97 \ \mu\mathrm{m}$. The movie is at $6 \times$ real time and has been JPEG compressed to reduce file size; reproduced by permission of The Royal Society of Chemistry \cite{Thijssen2010}.
\end{itemize}

\clearpage

% jtCHECK: OTR !!!
\section{Structure factors for image sequence in figure 3 (main text)}\label{sec:strucfac4imagseries}

% jtCHECK: OTR !!!
\begin{figure}[!h]
\begin{center}
\includegraphics[width=0.99\textwidth]{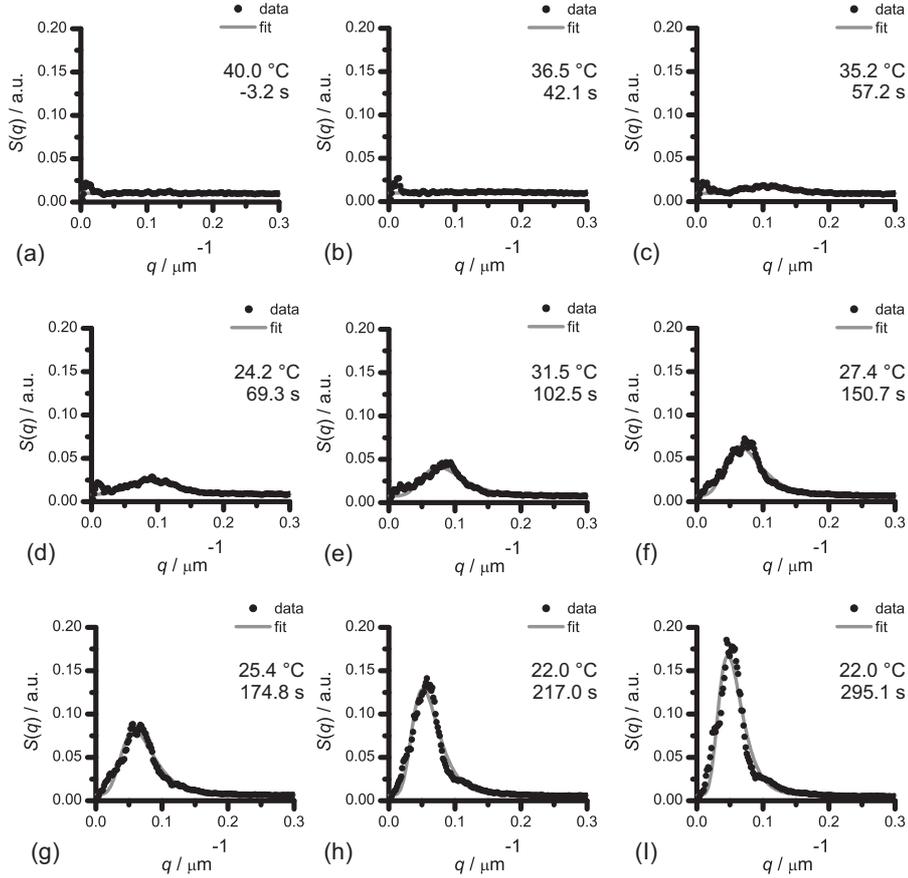}
\caption{Structure factors \Sq{} corresponding to figure 3(a-i) of the main text, which shows a confocal fluorescence image sequence of emulsification in a hexane-methanol mixture containing 2.0 vol\% of silica particles (\emph{Appendix \ref{sec:movieguide}}). The emergence and growth of the main(secondary) peak reflects the (subsequent) emergence of a dominant(secondary) length scale. Starting at $t = 0$ s, the sample was quenched from 40.0 \degC{} to 22.0 \degC{} at 5.0 \degCpm{}, resulting in the nucleation and growth of hexane-rich droplets in a methanol-rich continuous phase. The analysed images were recorded at 14 \micron{} from the top of the sample.}\label{fig:SI_Figure_Coalescence}
\end{center}
\end{figure}

\clearpage

%\bibliographystyle{unsrt}
%\bibliography{Thijssen_2010_EQA_References_DATA}

\end{document}